\documentclass[aps,twocolumn,superscriptaddress,groupedaddress]{revtex4-1}  

\usepackage{graphicx}  
\usepackage{dcolumn}   
\usepackage{bm}        
\usepackage{amssymb}   
\usepackage{siunitx}
\usepackage[english]{babel} 
\usepackage{blindtext}
\usepackage{inputenc}
\usepackage{amsmath}
\usepackage{amssymb}
\usepackage{graphicx}
\usepackage{natbib}
\usepackage{lipsum}
\usepackage{mathrsfs}
\usepackage{color}
\usepackage[abs]{overpic}

\begin{document}
\title{Tunnelling of transverse acoustic waves on a silicon chip}

\author{Nicolas P. Mauranyapin}
\thanks{These authors contributed equally to this work}
\author{Erick Romero}
\thanks{These authors contributed equally to this work}
\author{Rachpon Kalra}
\author{Glen Harris}
\author{Christopher G. Baker}
\author{Warwick P. Bowen}
\affiliation{ARC Centre for Engineered Quantum Systems, School of Mathematics and Physics, The University of Queensland, Brisbane, Queensland 4072, Australia}

\begin{abstract}
Nanomechanical circuits for transverse acoustic waves promise to enable new approaches to computing, precision biochemical sensing and many other applications. However, progress is hampered by the lack of precise control of the coupling between nanomechanical elements. Here, we demonstrate virtual-phonon coupling between transverse mechanical elements, exploiting tunnelling through a zero-mode acoustic barrier. This allows the construction of large-scale nanomechanical circuits on a silicon chip, for which we develop a new scalable fabrication technique. As example applications, we build mode-selective acoustic mirrors with controllable reflectivity and demonstrate acoustic spatial mode filtering. Our work paves the way towards applications such as fully nanomechanical computer processors and distributed nanomechanical sensors, and to explore the rich landscape of nonlinear nanomechanical dynamics.
\end{abstract}

\maketitle

\section{Introduction}

Large-scale nanomechanical circuits promise diverse applications, from heat mitigation in next-generation computer architectures~\cite{biswas2012high}, to integrated nanomechanical sensor arrays for biomedical diagnostics~\cite{baller2000cantilever}, nanomechanical computers robust to ionising radiation~\cite{furano2013review} and quantum information processing and storage technologies~\cite{pechal2018superconducting,bienfait2019phonon,satzinger2018quantum,shin2015control,fu2019phononic}. Transverse nanomechanical circuits are particularly important. They dominate applications in nanomechanical sensing and computing~\cite{longo2013rapid}, and in nonlinear dynamics~\cite{kurosu_-chip_2018}, due to the orders-of-magnitude higher sensitivity and nonlinearity that can be  provided by transverse acoustic waves (Supplemental Material \cite{Supp}). However, progress has been slowed by a lack of effective means to couple acoustic energy between transverse nanomechanical elements. As a result, most transverse nanomechanical circuits created to-date have employed only a few elements~\cite{wenzler2014nanomechanical}.

\begin{figure}[h!]
\begin{center}
\includegraphics[width=0.8\columnwidth]{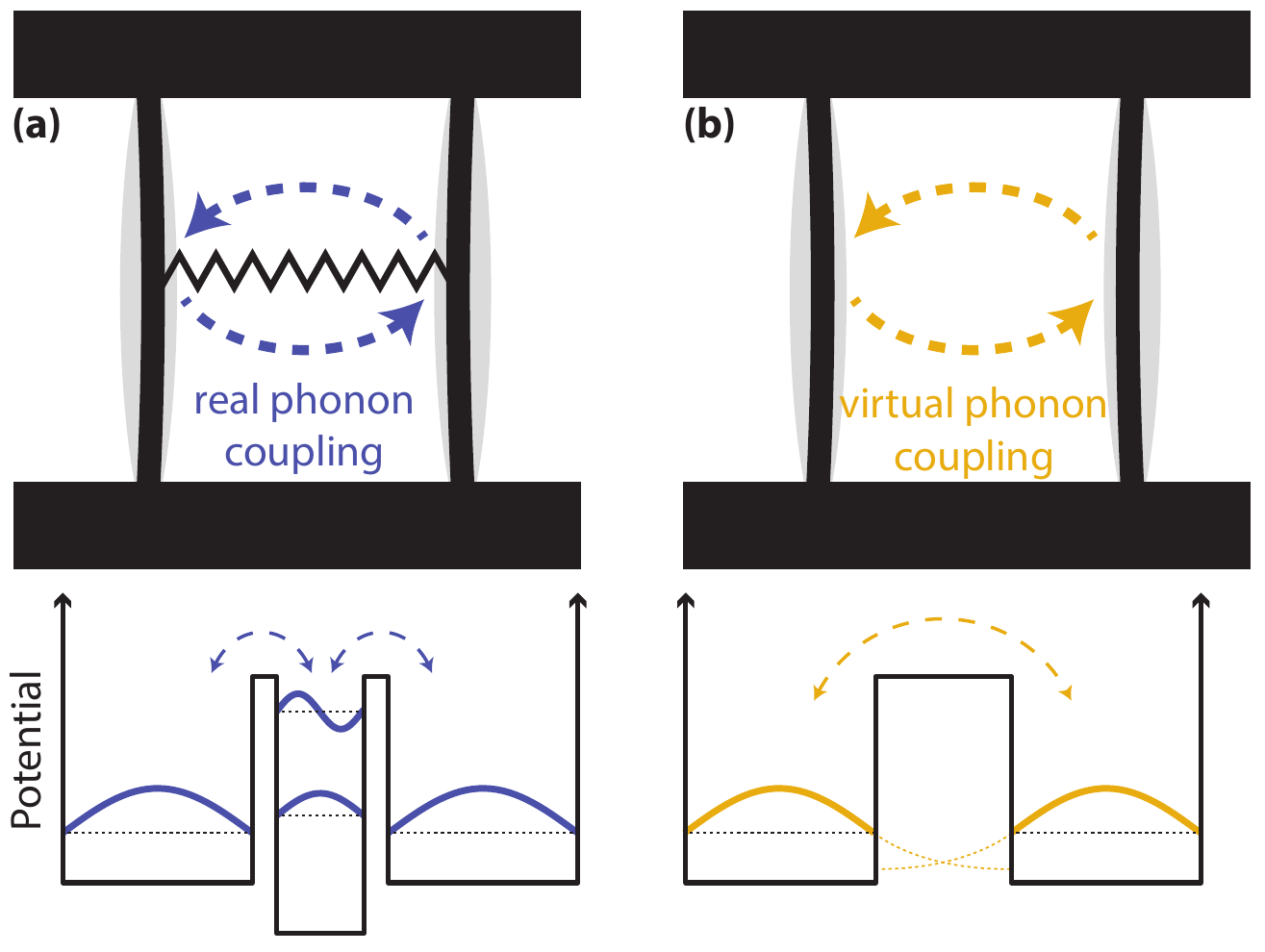}
\end{center}
\caption{Coupling types. (a) Spring-coupling. (b) Virtual phonon coupling. Top, schematics of two mechanical elements represented by doubled clamped beams coupled via real (case (a)) or virtual (case (b)) phonons. Bottom, corresponding potential energy landscape as function of position. In case (a), real phonons couple in and out of the spring via allowed states. However, in case (b), the absence of modes results in acoustic tunnelling by virtual phonons.
}
\label{fig:RealVirtualcoupling}
\end{figure}

Transverse nanomechanical coupling is conventionally achieved via an effective spring interaction, as illustrated in figure \ref{fig:RealVirtualcoupling}(a). For instance, pairs of nanomechanical resonators can be spring-coupled directly using a tether~\cite{lee2004mechanically,demirci2006mechanically,greywall2002coupled}, or indirectly through their mutual interaction with a common substrate~\cite{doster2019collective}. Acoustic phonons then couple from one element to the next through the intermediary spring. However, resonances in the spring capture acoustic energy and create complex frequency dependence, while  the long range of the coupling necessitates large device footprints that are impractical for many applications (see Supplemental Material \cite{Supp}).

In this paper, we implement a fundamentally different form of transverse-wave coupling, where energy is transferred by virtual phonons (figure \ref{fig:RealVirtualcoupling}(b)). To achieve this we develop a new architecture for nanomechanical circuitry based on sequences of zero-mode, single-mode and multimode acoustic waveguides. We show that transverse acoustic waves tunnel through the zero-mode waveguides, which act as tunnel barriers in which only virtual phonons can exist. The tunnel barriers exhibit no resonances due to the absence of real phonons, while the exponential decay of the acoustic excitation within them allows compact device footprints.

Our architecture is CMOS-compatible, allowing the construction of complex nanomechanical devices from a pattern of sub-wavelength-scale holes in a thin membrane. As example applications, we demonstrate mode-selective acoustic mirrors and spatial mode filters on a silicon chip -- capabilities required for spatial mode multiplexing and mode-cleaning in phononic circuits. Together, our results provide a pathway for scalable transverse nanomechanical circuitry, with broad applications from distributed sensing to nonlinear phononics.

\begin{figure*}[ht!]
\includegraphics[width=1.6\columnwidth]{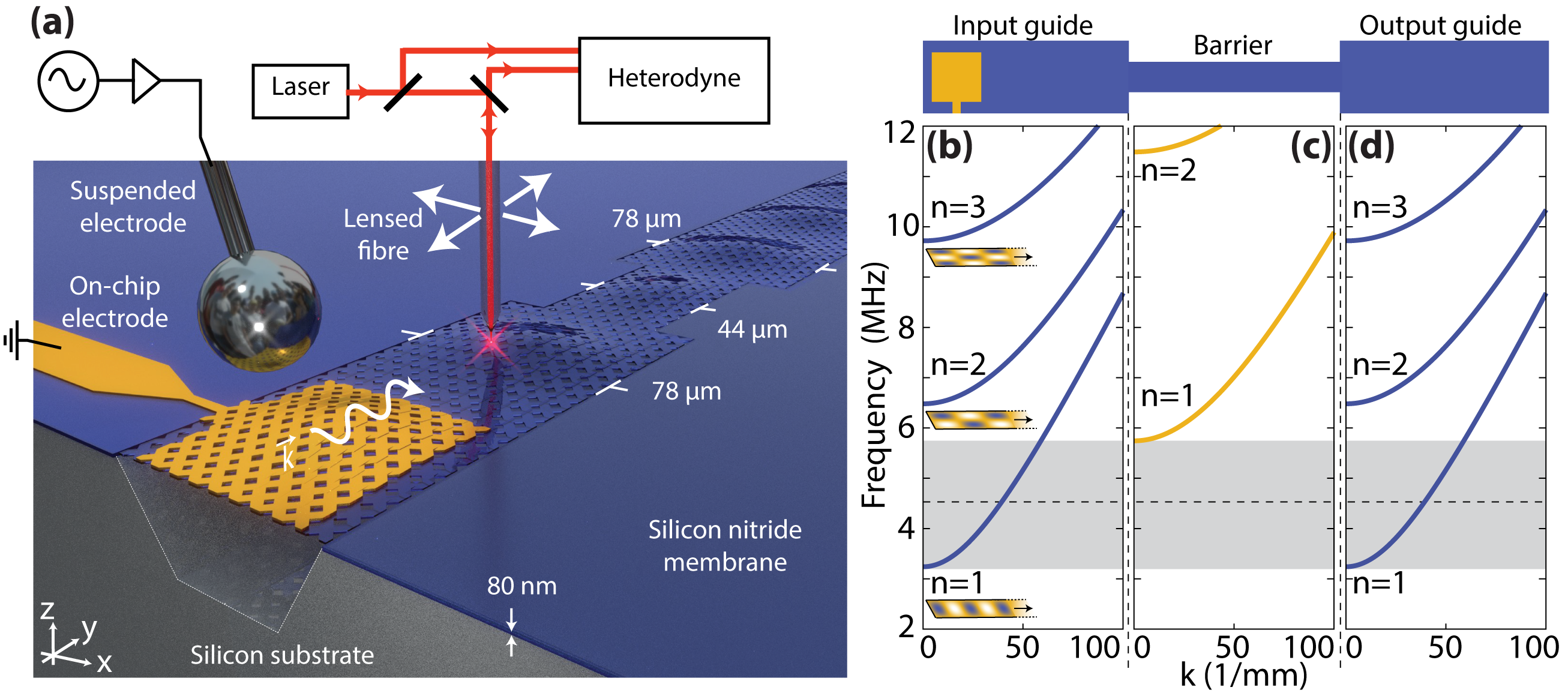}
\caption{Experimental setup and dispersion relation.  (a) The phononic device consists of an input waveguide, a tunnel barrier and an output waveguide made out of a $\sim80$~nm thick released  silicon nitride membrane. Mechanical waves are excited via electrostatic forces between a suspended and an on-chip electrode separated by $\sim 2 \mu$m (left). The mechanical vibrations are detected using heterodyne interferometry (right).  
(b)-(d) Phononic dispersion relation of the input waveguide, barrier and output waveguide respectively,  calculated from equation~\ref{eqn:Dispersion_Relation}. Mode profiles calculated from the solutions of the wave equation are shown as inset images in figure (b) for the first three modes ($n=1,2,3$). The grey shaded region in figure (b), (c) and (d) indicates the frequency band within the single-mode tunnelling regime. 
}
\label{fig:setup}
\end{figure*}

\section{Theory}\label{sec:Theory}

The devices developed here are made from a thin ($\sim80$~nm), highly stressed, silicon nitride (Si$_3$N$_4$) membrane grown upon a silicon (Si) wafer. Once released from the silicon, these membranes support acoustic waves with motion predominantly in the out-of-plane direction \cite{romero_propagation_2019} ($z$-direction  in figure \ref{fig:setup}(a)).  The membrane motion obeys a standard two-dimensional wave equation (see ref \cite{romero_propagation_2019})
with dispersion relation 
\begin{equation}\label{eqn:Dispersion_Relation}
\Omega=\sqrt{\frac{\sigma}{\rho}}\sqrt{ k_{y}^2 + \left(\frac{n \pi}{L_x}\right)^2 },
\end{equation}
where $\Omega$ is the angular frequency of the acoustic wave excitation, $\sigma$ the tensile stress of the membrane, $\rho$ the density of the membrane material, $k_y$ the wave number in the direction of propagation, $L_x$ the width of the waveguide in the transverse direction and $n$ an integer representing the transverse mode number. From this equation, one can see that each mode has a cut off frequency,
\begin{equation}
 \Omega_{c,n}=\sqrt{\frac{\sigma}{\rho}}\left(\frac{n \pi}{L_x}\right)
 \label{eqn:CutOffFreq}
 \end{equation}
below which the wave number $k_y$ is imaginary. 
 Therefore if the mode $n$ is excited at a frequency below $\Omega_{c,n}$, there are no real wave vector available in the dispersion relation, the amplitude of the acoustic wave will decay exponentially over distance. Such evanescent waves are associated to the presence of virtual phonons by opposition to real phonons which are associated to propagating acoustic waves (see figure~\ref{fig:RealVirtualcoupling}).
Thus,  there will be a range of frequencies ($\Omega < \Omega_{c,1}$) where no mode can propagate, followed by a range of frequencies ($\Omega_{c,1} < \Omega < \Omega_{c,2}$) where only the first  transverse mode is allowed and the system is single-mode  \cite{romero_propagation_2019}. This is the physical phenomenon which will be leveraged in this paper to engineer acoustic tunnelling. Above the second transverse mode cut off frequency, $ \Omega > \Omega_{c,2}$, the membrane can support several acoustic modes and becomes multi-mode.

We designed phononic devices, as shown in Figure \ref{fig:setup}(a), composed of an input waveguide connected to a narrower-width waveguide section (referred to as the ``tunnel barrier'' henceforth), itself connected to an output waveguide of the same width as the input waveguide. Since the dispersion relation depends on the width of the waveguide, the input and output waveguides have different dispersion relations  to that of  the tunnel barrier, as displayed in figure \ref{fig:setup}(b)-(d). This leads to different first mode cut off frequencies ($\Omega_{c,1}^{\rm{guides}}$ and $\Omega_{c,1}^{\rm{barrier}}$), and to different operating regimes depending on the excitation frequency $\Omega$.

If $\Omega_{c,1}^{\rm{guide}} < \Omega < \Omega_{c,1}^{\rm{barrier}}$ the acoustic wave can propagate in the guides via its first transverse mode but decays exponentially in the tunnel barrier, partially reflecting and partially tunnelling into the output waveguide. 
We henceforth refer to this range of frequencies, where the device acts as an acoustic mirror with controllable reflectivity, as the ``single-mode tunnelling regime''.
The ratio of reflection to tunnelling depends on the magnitude of the exponential decay, and therefore on both the length of the tunnel barrier and the amplitude exponential decay length $\gamma$. For frequencies within the single-mode tunnelling regime, $\gamma$ is given by: 
\begin{equation}\label{eqn:decay_rate}
\gamma=\left(\left(\frac{\pi}{L_x}\right)^2 - \Omega^2\frac{\rho}{\sigma} \right)^{-1/2} .
\end{equation}
Thus, the strength of the coupling between the two waveguides can be engineered by carefully choosing the length of the tunnel barrier or the driving frequency. The closer the frequency to $\Omega_{c,1}^{\rm{barrier}}$, the stronger the coupling is. Indeed, as $\Omega_{c,1}^{\rm{barrier}}$ is approached the decay length of the wave approaches infinity  so that the acoustic wave is barely attenuated. On the other hand, at frequencies close to $\Omega_{c,1}^{\rm{guide}}$ the decay length reaches a minimum of  $\left((\pi/L_x^{\rm{barrier}})^2-(\pi/L_x^{\rm{guide}})^2\right)^{-1/2}$, leading to maximal attenuation.

\section{Meshed Mechanical Systems}
\label{sec:Meshed_Mechanical_Systems}

Previously, processes used to fabricate membrane-based phononic devices have relied either on deep-backside etching \cite{romero_propagation_2019,
romero_engineering_2020} or on the use of wavelength-scale holes in the membrane to enable front-side etching \cite{hatanaka2014phonon,cha_experimental_2018}. For backside etching processes, the need to etch through a several hundred micron-thick substrate has limited both precision and feature size of the phononic components, while in the case of front-side etching, the wavelength-scale hole pattern \cite{kurosu_-chip_2018,
kurosu_mechanical_2020,
cha_electrical_2018} directly affects the dispersion relations of the guided modes, and limits the ability to fabricate arbitrary waveguide shapes.

\begin{figure*}[ht!]
\includegraphics[width=1.6\columnwidth]{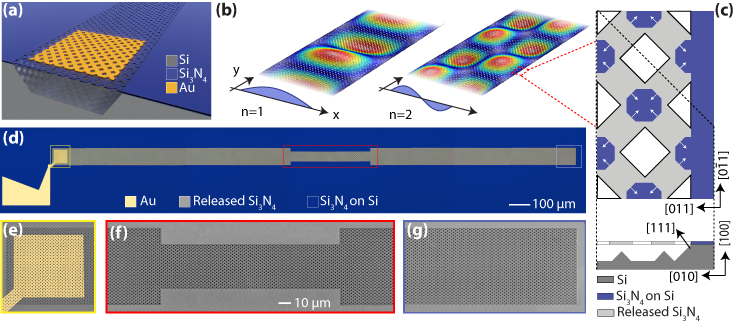}
\caption{Meshed phononic waveguide fabrication. (a) Schematic of the meshed silicon nitride membrane released from the silicon substrate and gold actuation electrode. (b) Finite element simulations showing the transverse mode profile of the first ($n=1$) and second ($n=2$) acoustic modes of a meshed waveguide. 
(c) Schematic of a snapshot of the chip during the KOH etch. The top figure is a top-view of the chip and the mesh pattern in shown in white, the released silicon nitride is displayed in light grey and the non-released membrane in blue. The bottom figure is a cut of the top figure along the black dashed line. (d) Optical microscope image showing the input waveguide with a meshed gold electrode (yellow square), the tunnel barrier (red rectangle) and the output waveguide (blue rectangle). (e)-(g) False color scanning electron micrograph of the actuation region with a gold electrode, the tunnel barrier and the end of the output waveguide, respectively.}
\label{fig:Modes_Mesh}
\end{figure*}

Here, we develop a fabrication platform that overcomes these two issues. It is based on a far sub-wavelength hole pattern resulting in a ``meshed'' silicon nitride membrane through which the underlying silicon substrate can be etched away from the front-side. The pattern consists of square holes of 1~$\mu$m by 1~$\mu$m periodically separated (center to center) by 3~$\mu$m (see figure \ref{fig:Modes_Mesh}(a)). These lengths are approximately two orders of magnitude smaller than the typical wavelengths of the guided acoustic waves. Therefore, the interaction of the supported acoustic waves with the holes is expected to be greatly suppressed, leaving the dispersion relation equation (\ref{eqn:Dispersion_Relation}) essentially unaffected, with only  the ratio $\sqrt{\sigma/\rho}$ reduced by 12\% compared to a non-patterned membrane (see appendix \ref{sec:MeshMembraneParameters}). This is verified through finite element simulations of the $n=1$ and $n=2$ transverse modes of a meshed membranes (displayed in figure \ref{fig:Modes_Mesh}(b)), which show no apparent differences in mode shape from the corresponding acoustic waves propagating on a non-patterned membrane. 


The meshed phononic devices are fabricated on a chip from a commercial wafer with an $\sim80$~nm stoichiometric Si$_3$N$_4$ film (LPCVD deposition, initial tensile stress $\sigma_0 = 1$~GPa) deposited on a silicon substrate. The hole pattern in the  silicon nitride membrane is defined through a combination of  electron beam lithography and reactive ion etching. 
The devices are released through anisotropic wet etching of the underlying silicon using a potassium hydroxide (KOH) solution. Aligning the waveguides along the [011] crystal axis results in near atomically smooth side-walls (see figure \ref{fig:Modes_Mesh}(c) and appendix \ref{sec:MeshMembraneFabricationProcess}).

 An optical image of the phononic device is shown in figure~\ref{fig:Modes_Mesh}(d), with the colors of different layers naturally arising due to the thin film interference effects.
The yellow, red and blue frames respectively enclose the on-chip electrode, the tunnel barrier and the end of the output waveguide. Scanning electron images of these regions are shown in figure~\ref{fig:Modes_Mesh}(e)-(g) respectively. One can see the straight planar side-walls closely following the pattern of holes, and that features smaller than 10~$\mu$m can be achieved in the released membrane using this technique.
Despite the high initial tensile stress, the process is remarkably robust, generally achieving yields of 100~\% for chips containing as many as 48 devices.


\section{Setup}
\label{sec:setup}

We fabricated devices with input and output waveguide widths of $L_x = 78$~$\mu$m and a tunnel barrier width of $L_x = 44$~$\mu$m. The input and output waveguides are $\sim 1$~mm long in the direction of propagation  and different tunnel barrier lengths are investigated. Given the choice of widths, the cut-off frequency of the first mode of the guides and the tunnel barrier are found from equation (\ref{eqn:CutOffFreq}) to be $\Omega_{c,1}^{\rm{guide}}/2\pi = 3.2$~MHz and $\Omega_{c,1}^{\rm{barrier}}/2\pi = 5.6$~MHz, respectively. This provides a frequency band of 2.4~MHz (grey shading in figure \ref{fig:setup}(b)-(d)) for which single-mode acoustic tunnelling can be investigated.

Acoustic waves are launched into the device through electrostatic actuation \cite{romero_propagation_2019} between a gold electrode patterned on the input waveguide and a suspended electrode, as shown in figure \ref{fig:setup}(a). To detect the motion of the membrane, we use an optical lensed fibre combined with heterodyne detection 
(see Refs.~\cite{romero_propagation_2019,mauranyapin_evanescent_2017} for more details). 
Once detected, the heterodyne signal produces a photocurrent whose amplitude is proportional to the amplitude of the membrane motion at the point of focus of the lensed fibre. Therefore, by scanning the lensed fibre across and along the device, we can determine the amplitude of an acoustic wave at any position.

\section{Results}

\subsection{Exponential decay}

\begin{figure*}[ht!]
\includegraphics[width=1.6\columnwidth]{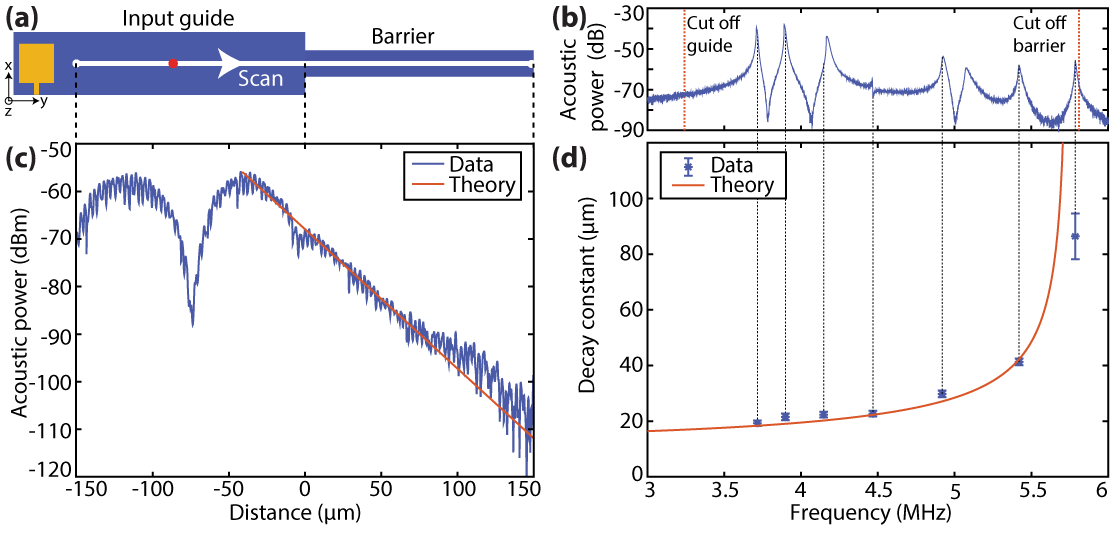}
\caption{Exponential decay. (a) Scheme of one-dimensional scan. The lensed fibre is scanned from the input waveguide until the end of the tunnel barrier at $x= 0.5L_x$ at a rate of 10 steps per second with a step size varying from scan to scan between 880 to 900 nm. The amplitude of the mechanical signal is recorded on a spectrum analyser at 0 span and 10 Hz resolution bandwidth.  
(b) Network analysis performed in the middle of the input waveguide ($x= 0.5L_x$ and $y= 0.5L_y$ represented by the red dot in (a)). The cut off frequency of the waveguide and tunnel barrier first mode are displayed in red and calculated using the equations from the theoretical discussed previously.
(c) Experimental scan of the input waveguide and tunnel barrier (150~$\mu$m long). The power of the acoustic wave is displayed versus the scanning distance for a typical scan at a driving frequency of 4.9~MHz. The fast periodic modulations on the trace originates from the mesh pattering of the waveguide, which modulates the reflection of the probing optical beam. Theoretical prediction in red is calculated from the amplitude of the wave at the guide/tunnel barrier interface (distance axis = 0 $\mu$m) with no fitting parameter.
 (d) Extracted evanescent decay constant versus drive frequency measured at the points in (b) indicated by the dashed lines. The error bars are calculated from the standard deviation over 6 different scans for each frequency. The red theoretical line is obtained from equation \ref{eqn:decay_rate} with no fitting parameter.}
\label{fig:ExpDecay}
\end{figure*}

We firstly investigate how the acoustic wave decays in the tunnel barrier. To do this, we use a device with a $150$~$\mu$m long tunnel barrier, significantly longer than the typical acoustic wave decay length in the barrier. 
To investigate the response of the device, we perform a network analysis with the lensed fibre placed in the middle (both in the $x$ and $y$-directions) of the input waveguide (see figure \ref{fig:ExpDecay}(a) red dot). The results are displayed in figure \ref{fig:ExpDecay}(b) and we observe no response from the device at frequencies below  $\sim3.5$~MHz, consistent with the theoretical cut-off frequency of the waveguide of 3.2~MHz. Above this frequency, we observe a series of resonant peaks. 
These peaks are expected due to the finite dimensions of the device, with impedance mismatch between  the released silicon nitride membrane and the silicon substrate causing reflection of the acoustic wave at each end of the input waveguide \cite{romero_propagation_2019}.

The quality factors of the observed resonances can be used to provide an upper bound on the losses of the acoustic wave during propagation \cite{romero_propagation_2019}. We observe quality factors as high as 5,000. This corresponds to a loss per unit length as low as 0.4~dB~cm$^{-1}$. To our knowledge, this is the lowest propagation loss achieved for megahertz frequencies  in a phononic waveguide at room temperature \cite{romero_propagation_2019,fu2019phononic}. This indicates an absence of additional damping introduced by the sub-wavelength mesh used for fabrication.

We use the amplitude enhancement provided by these resonances to investigate how the acoustic wave decays in the tunnel barrier. To do so, the device is  continuously  driven at one of the resonance frequencies within the single-mode tunnelling regime while the lensed fibre is scanned in the $y$-direction along the input waveguide and the tunnel barrier at the $x$-position $x = L_x/2$ (see figure \ref{fig:ExpDecay}(a)). The amplitude of the membrane motion is recorded continuously over the scanning distance.
 
A typical scan is shown in figure \ref{fig:ExpDecay}(c) for a driving frequency of 4.9~MHz. The power of the acoustic wave is plotted versus the distance scanned by the fibre along the direction of propagation. The coordinate $y=0$ corresponds to the junction between the input waveguide and the tunnel barrier. One can see that, for $y>0$, the acoustic wave amplitude decays exponentially in the tunnel barrier as expected for this frequency. For $y<0$, we observe standing wave oscillations expected due to the resonant nature of the input waveguide. 
The red line in Figure \ref{fig:ExpDecay}(c) corresponds to the theoretical decay expected for an acoustic wave at frequency 4.9 MHz. The good agreement with the experimental data verifies that the simple theoretical model discussed previously is appropriate for our experimentally fabricated devices.

To map the dependence of $\gamma$ with the drive frequency, this process was repeated for seven different drive frequencies (corresponding to the peaks shown in figure \ref{fig:ExpDecay}(b)).
The results are shown in figure \ref{fig:ExpDecay}(d) and compared to the theoretical prediction of equation (\ref{eqn:decay_rate}) without any fitting parameters. 
As expected, the decay length increases with frequency and tends to infinity as the cut-off frequency of the first mode of the tunnel barrier is approached. This experimentally demonstrates that varying the drive frequency provides the ability to tune the decay length by more than a factor of four. The good agreement with theory demonstrates that the phononic decay can be precisely and reliably engineered,  opening a path toward scalable phononic circuitry.

\subsection{Imaging acoustic tunnelling}

\begin{figure*}[ht!]
\includegraphics[width=1.6\columnwidth]{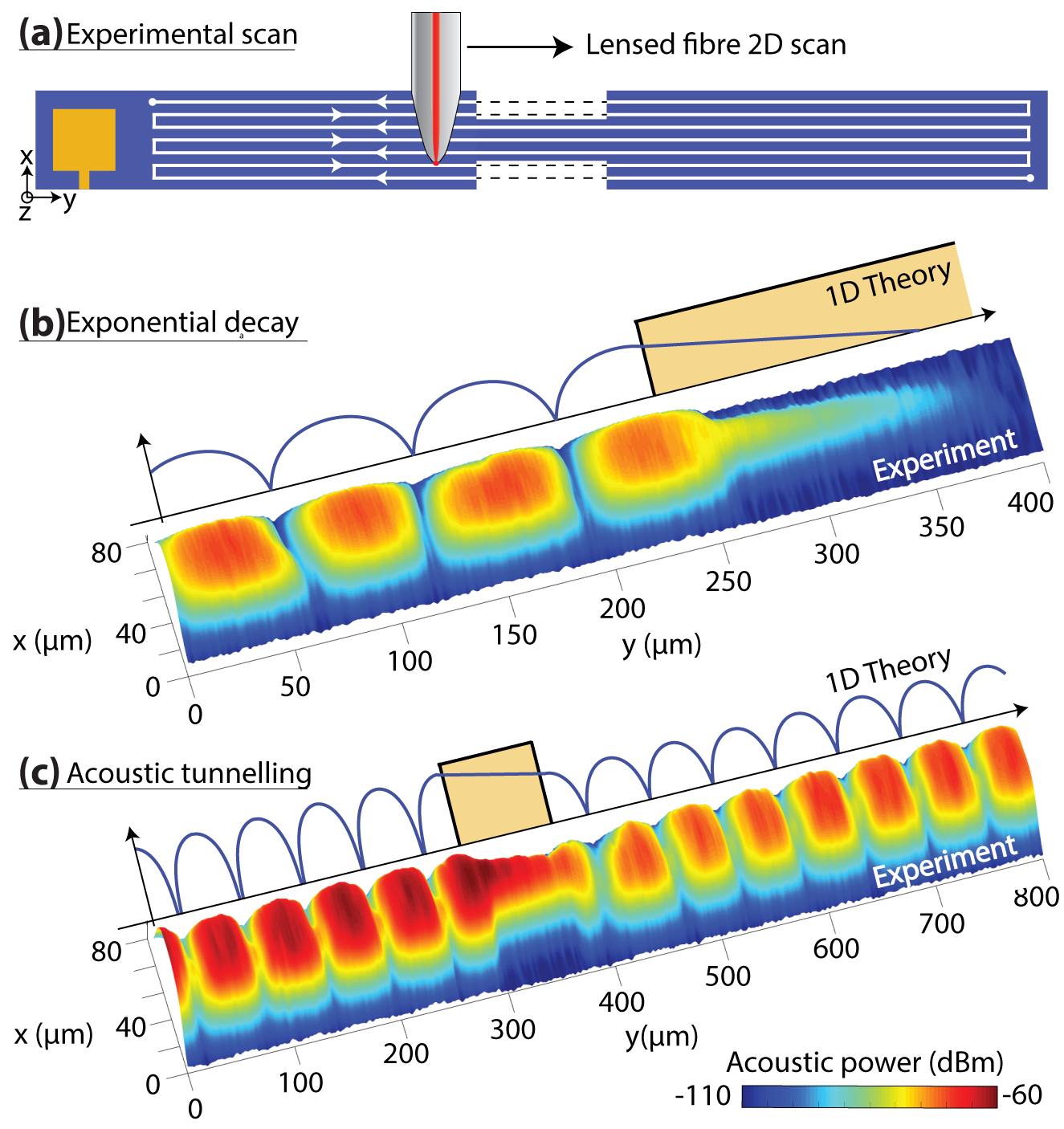}
\caption{Imaging of acoustic decay and tunnelling. (a) Schematic of the raster scan imaging process where the lensed fibre is scanned in the $y$-direction for several $x$-positions. 
The different scans for all the $x$-positions are assembled in post processing.  (b) Imaging of  acoustic decay in a device with a 150~$\mu$m long tunnel barrier. The device was continuously driven at 4.9~MHz while imaging. Top schematics show a theoretical prediction of acoustic exponential decay in one dimension. (c) Imaging of acoustic tunnelling. The device was driven at 5.5 MHz and has a 75~$\mu$m long tunnel barrier. 
The experimental data in (b) and (c) was smoothed with a Gaussian filter.}\label{fig:Tunnelling}
\end{figure*}

By raster-scanning the lensed fibre in both $x$ and $y$-direction as shown in figure \ref{fig:Tunnelling}(a), two-dimensional images of the acoustic wave are recorded. Figures \ref{fig:Tunnelling}(b) and (c) display two-dimensional scans of two different devices. In figure \ref{fig:Tunnelling}(b), the device has a tunnel barrier of 150~$\mu$m length and the image is recorded for a driving frequency of 5.4~MHz. As discussed previously but now in two dimensions, we observe a resonant wave in the input waveguide which exponentially decays below the noise floor in the tunnel barrier with the acoustic wave fully reflected. This is expected at this frequency because the decay length of the wave is around 42~$\mu$m (see figure \ref{fig:ExpDecay}(d)) which is three times smaller than the tunnel barrier length.

The second device (figure \ref{fig:Tunnelling}(c)) has a shorter 75~$\mu$m long tunnel barrier and is driven at a frequency of  5.5~MHz. We observe resonance in the input waveguide, then a short exponential decay in the tunnel barrier followed by resonant build up again in the output waveguide. 
Over all, we observe transmission through the tunnel barrier of 10\%  via virtual phonon coupling, similar to the frustrated total internal reflection observed between two prisms in photonics \cite{stahlhofen2006evanescent}.

\subsection{Acoustic mode filtering}
\label{sec:Acoustic_filtering}

\begin{figure*}[ht!]
\includegraphics[width=1.6\columnwidth]{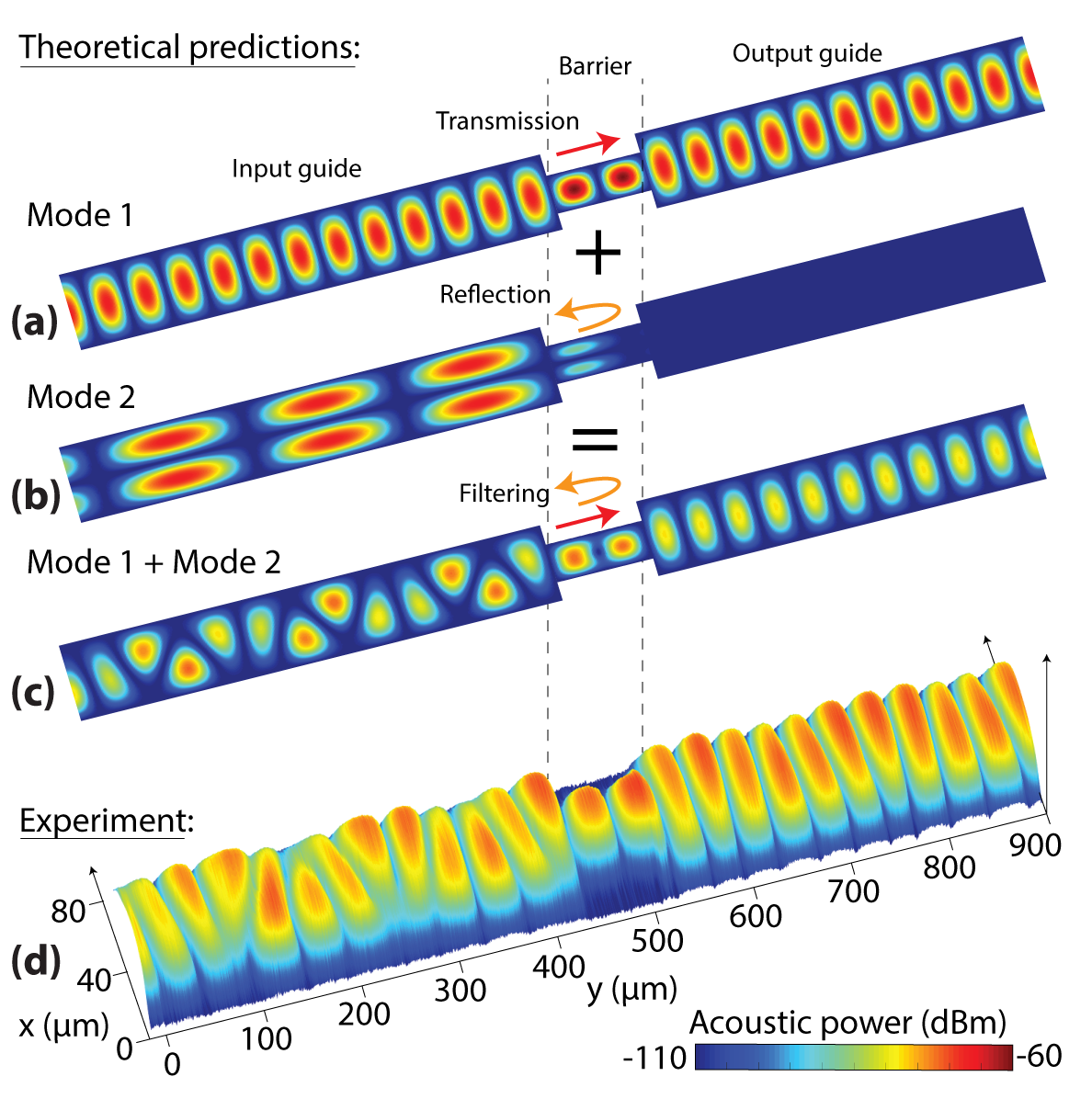}
\caption{Acoustic mode filtering. (a)-(c) Theoretical predictions of the acoustic power of a device driven at a frequency between the guides second and third mode cut off frequency ($\Omega_{c,2}^{guide} < \Omega < \Omega_{c,3}^{guide}$). (a) Acoustic power if only the first mode is excited and (b) if only the second mode is excited for the same frequency. (c) corresponds to the interference pattern of the first two modes with identical amplitude and is calculated by summing (a) and (b) over the entire device. (d) Experimental imaging of a device with a 75~$\mu$m long tunnel barrier driven at 8.1 MHz experimentally showing acoustic mode filtering. We believe that both modes can be excited simultaneously in the input waveguide because the on-chip electrode is not symmetric in the propagation direction. Note that in this configuration, due to resonances in the output waveguide,  it is possible to selectively enhance the second mode amplitude by driving at its resonance frequency even if it is greatly suppressed by the tunnel barrier. The observed second mode amplitude in the output waveguide will be affected by this resonances and, even though filtering is clearly evident, prevent an exact estimation of the filtering efficiency. The experimental data was smoothed with a Gaussian filter.}
\label{fig:ModeFiltering}
\end{figure*}

To illustrate one application of the ability to engineer acoustic wave tunnelling, we consider acoustic spatial mode filtering. Spatial mode filtering is an important capability for phononic circuitry. Similarly to photonics, it allows spatial mode multiplexing, control of spatial dispersion, and filtering of scattering from defects \cite{luo2014wdm}, among other prospective applications. 
Here, we demonstrate it by arranging a scenario where both guides can support the first two transverse modes but the tunnel barrier remains in the single-mode regime. In this case, if both modes are excited in the input waveguide, only the first transverse mode will be allowed to transmit through the barrier into the output waveguide, with the second mode fully reflected.

This regime can be achieved by driving the device at frequencies below the second mode cut off frequency of the barrier $\Omega_{c,2}^{\rm{barrier}}/2\pi = 11.5$~MHz and between the  second and third mode cut off frequencies of the waveguides, $\Omega_{c,2}^{\rm{guide}}/2\pi = 6.5$~MHz and  $\Omega_{c,3}^{\rm{guide}}/2\pi = 9.7$~MHz respectively. Figure  \ref{fig:ModeFiltering}(a) and b show theoretical predictions of the  propagation of the first and second acoustic modes in the waveguides, respectively. 
If both modes are driven simultaneously, they will interfere creating spatial patterns such as shown in figure \ref{fig:ModeFiltering}(c).
Choosing a 8.1~MHz excitation frequency, the decay length of the second mode is estimated, (using equation \ref{eqn:decay_rate} adapted for the second transverse mode), to be around 7~$\mu$m. After a 75~$\mu$m long tunnel barrier, this is predicted to exponentially reduce the power in the second mode by a factor of $2\times10^9$ (or -93~dB) .
The first mode, on the other hand, passes essentially unaffected. We then expect that the second transverse mode will be spatially filtered by the device. Figure \ref{fig:ModeFiltering}(d) shows an experimental image of the acoustic propagation in this configuration. The results are consistent with the theory, showing clear acoustic mode filtering.

\section{Discussion}

Transverse nanomechanical circuits have many potential applications~\cite{biswas2012high,baller2000cantilever,furano2013review,pechal2018superconducting,bienfait2019phonon,satzinger2018quantum,shin2015control,fu2019phononic}. However, the inability to effectively couple transverse nanomechanical elements has slowed progress, limiting circuits to simple configurations of a few components~\cite{wenzler2014nanomechanical}. Our work provides a resolution to this issue, introducing a new coupling method based on acoustic tunnelling and demonstrating that it can be used to construct mode-selective acoustic mirrors and  acoustic mode filters. The absence of propagating modes within the coupler removes the resonances intrinsic to other methods~\cite{fang2016optical} (see Supplemental Material \cite{Supp}). This greatly suppresses energy capture inside the coupler and significantly increases tolerance to fabrication imperfections. 
 Tunnelling also greatly reduces the length scale over which coupling occurs, providing the possibility of orders-of-magnitude reduced device footprints compared to other single layer devices (see Supplemental Material \cite{Supp}).


To observe tunneling we report the development of a scalable silicon-chip-based architecture for nanomechanical circuitry, which we construct using a new CMOS-compatible fabrication approach. The architecture is based on zero-mode tunnel barriers, within which only virtual phonons can exist, integrated with single- and multi-mode acoustic waveguides. This guide-barrier approach is analogous to evanescent coupling in optics~\cite{haus1991coupled} which has, for example, been used to build complex photonic circuits~\cite{tian2017experimental}, spatial filters~\cite{gabrielli2009silicon}, add-drop filters~\cite{qiang2007optical} and coupled resonators~\cite{bekker2018free}. As in photonics, the use of single-mode waveguides offers immunity to deleterious effects such as modal dispersion, spatial mode mismatch, and scattering from defects. 


Acoustic tunnelling was first demonstrated for longitudinal acoustic waves in the 1970s ~\cite{narayanamurti1979selective} and has recently been exploited to build basic nanomechanical circuits~\cite{shin2015control,fu2019phononic}.  Phonon dimers have also recently been reported \cite{catalini2020soft}. However, our work is the first to use transverse tunnelling to couple elements for nanomechanics circuitry. Transverse waves dominate applications in areas such as nanomechanical sensing, computing and nonlinear mechanics~\cite{longo2013rapid,lee2004mechanically}. For instance, they provide four orders-of-magnitude improved precision in nanomechanical mass sensing~\cite{taylor2012cavity} and two orders of magnitude higher mechanical nonlinearity (see Supplemental Material \cite{Supp}). We therefore expect the results presented here to open up diverse applications from distributed sensing, to quantum information and nanomechanical computing.


\section*{Acknowledgement}
This work was performed in part at the Queensland node of the Australian National Fabrication Facility. A company established under the National Collaborative Research Infrastructure Strategy to provide nano and microfabrication facilities for Australia's researchers.
The authors acknowledge the facilities, and the scientific and technical assistance, of the Australian Microscopy \& Microanalysis Research Facility at the Centre for Microscopy and Microanalysis, The University of Queensland.
The authors thank Elliot Cheng and Tihan Bekker for their help for the fabrication and their comments on the manuscript.
\textbf{Funding:} This research was primarily funded by the Australian Research Council and Lockheed Martin Corporation through the Australian Research Council Linkage Grant LP160101616. Support was also provided by the Australian Research Council Centre of  Excellence for Engineered Quantum Systems (CE170100009). R.K., C.G.B and W.P.B. acknowledge fellowships from the University of Queensland (UQFEL1719237) and the Australian Research Council (DE190100318 and FT140100650), respectively.

\appendix

\section{Mesh membrane parameters}
\label{sec:MeshMembraneParameters}

Compared to a typical non-patterned membrane, the mesh on the silicon nitride causes the  membrane to relax leading to an effective stress equal to $\sigma=\sigma_0(1-\nu)$ \cite{fedorov2019generalized}, where $\sigma_0 = 1$~GPa is the deposition tensile stress of the non-patterned silicon nitride and $\nu=0.22$ the Poisson's ratio of silicon nitride. The effective stress of the meshed membrane is therefore equal to $\sigma = 0.88$~GPa and its density is the density of the silicon nitride, $\rho=3200$~kg/m$^3$.

\section{Mesh membrane fabrication process}
\label{sec:MeshMembraneFabricationProcess}

The electrodes are patterned using electron beam lithography on a double layer of polymethyl methacrylate (PMMA) resist, followed by 50~nm of gold evaporation and lift-off. The mesh array is aligned to the gold electrodes and patterned using AR-P electron beam resist. The mesh is formed by etching the exposed Si$_3$N$_4$ film using reactive ion etching with a plasma of CHF$_3$ and SF$_6$. The AR-P resist is then stripped off with oxygen plasma. The underlying silicon is removed using a solution of low concentration potassium hydroxide (KOH) combined with isopropyl alcohol until the membrane is released as represented in figure \ref{fig:Modes_Mesh}(c). The chip is dried in a CO$_2$ critical point dryer. 

\section{Experimental parameters}
\label{ExperimentalParameters}

To mechanically drive the device a signal generator at frequency $\Omega/2\pi$ and 0 dBm power is connected to a 30 V DC supply, amplified by 25 dBm and sent to the suspended electrode $\sim 2 \mu$m above the gold on-chip electrode, to generate acoustic waves via electrostatic forces.

Experiments are performed in a high vacuum chamber (pressure $10^{-7}$ mbar) to eliminate any air damping of the membrane motion \cite{naesby_effects_2017}.

\bibliography{PRA_Library,Tunnelling_of_transverse_acoustic_waves_on_a_silicon_chip}

\providecommand{\noopsort}[1]{}\providecommand{\singleletter}[1]{#1}%
\begin{thebibliography}{36}%
\makeatletter
\providecommand \@ifxundefined [1]{%
 \@ifx{#1\undefined}
}%
\providecommand \@ifnum [1]{%
 \ifnum #1\expandafter \@firstoftwo
 \else \expandafter \@secondoftwo
 \fi
}%
\providecommand \@ifx [1]{%
 \ifx #1\expandafter \@firstoftwo
 \else \expandafter \@secondoftwo
 \fi
}%
\providecommand \natexlab [1]{#1}%
\providecommand \enquote  [1]{``#1''}%
\providecommand \bibnamefont  [1]{#1}%
\providecommand \bibfnamefont [1]{#1}%
\providecommand \citenamefont [1]{#1}%
\providecommand \href@noop [0]{\@secondoftwo}%
\providecommand \href [0]{\begingroup \@sanitize@url \@href}%
\providecommand \@href[1]{\@@startlink{#1}\@@href}%
\providecommand \@@href[1]{\endgroup#1\@@endlink}%
\providecommand \@sanitize@url [0]{\catcode `\\12\catcode `\$12\catcode
  `\&12\catcode `\#12\catcode `\^12\catcode `\_12\catcode `\%12\relax}%
\providecommand \@@startlink[1]{}%
\providecommand \@@endlink[0]{}%
\providecommand \url  [0]{\begingroup\@sanitize@url \@url }%
\providecommand \@url [1]{\endgroup\@href {#1}{\urlprefix }}%
\providecommand \urlprefix  [0]{URL }%
\providecommand \Eprint [0]{\href }%
\providecommand \doibase [0]{http://dx.doi.org/}%
\providecommand \selectlanguage [0]{\@gobble}%
\providecommand \bibinfo  [0]{\@secondoftwo}%
\providecommand \bibfield  [0]{\@secondoftwo}%
\providecommand \translation [1]{[#1]}%
\providecommand \BibitemOpen [0]{}%
\providecommand \bibitemStop [0]{}%
\providecommand \bibitemNoStop [0]{.\EOS\space}%
\providecommand \EOS [0]{\spacefactor3000\relax}%
\providecommand \BibitemShut  [1]{\csname bibitem#1\endcsname}%
\let\auto@bib@innerbib\@empty
\bibitem [{\citenamefont {Biswas}\ \emph {et~al.}(2012)\citenamefont {Biswas},
  \citenamefont {He}, \citenamefont {Blum}, \citenamefont {Wu}, \citenamefont
  {Hogan}, \citenamefont {Seidman}, \citenamefont {Dravid},\ and\ \citenamefont
  {Kanatzidis}}]{biswas2012high}%
  \BibitemOpen
  \bibfield  {author} {\bibinfo {author} {\bibfnamefont {K.}~\bibnamefont
  {Biswas}}, \bibinfo {author} {\bibfnamefont {J.}~\bibnamefont {He}}, \bibinfo
  {author} {\bibfnamefont {I.~D.}\ \bibnamefont {Blum}}, \bibinfo {author}
  {\bibfnamefont {C.-I.}\ \bibnamefont {Wu}}, \bibinfo {author} {\bibfnamefont
  {T.~P.}\ \bibnamefont {Hogan}}, \bibinfo {author} {\bibfnamefont {D.~N.}\
  \bibnamefont {Seidman}}, \bibinfo {author} {\bibfnamefont {V.~P.}\
  \bibnamefont {Dravid}}, \ and\ \bibinfo {author} {\bibfnamefont {M.~G.}\
  \bibnamefont {Kanatzidis}},\ }\href@noop {} {\bibfield  {journal} {\bibinfo
  {journal} {Nature}\ }\textbf {\bibinfo {volume} {{489}}},\ \bibinfo {pages}
  {414} (\bibinfo {year} {2012})}\BibitemShut {NoStop}%
\bibitem [{\citenamefont {Baller}\ \emph {et~al.}(2000)\citenamefont {Baller},
  \citenamefont {Lang}, \citenamefont {Fritz}, \citenamefont {Gerber},
  \citenamefont {Gimzewski}, \citenamefont {Drechsler}, \citenamefont
  {Rothuizen}, \citenamefont {Despont}, \citenamefont {Vettiger}, \citenamefont
  {Battiston} \emph {et~al.}}]{baller2000cantilever}%
  \BibitemOpen
  \bibfield  {author} {\bibinfo {author} {\bibfnamefont {M.~K.}\ \bibnamefont
  {Baller}}, \bibinfo {author} {\bibfnamefont {H.~P.}\ \bibnamefont {Lang}},
  \bibinfo {author} {\bibfnamefont {J.}~\bibnamefont {Fritz}}, \bibinfo
  {author} {\bibfnamefont {C.}~\bibnamefont {Gerber}}, \bibinfo {author}
  {\bibfnamefont {J.~K.}\ \bibnamefont {Gimzewski}}, \bibinfo {author}
  {\bibfnamefont {U.}~\bibnamefont {Drechsler}}, \bibinfo {author}
  {\bibfnamefont {H.}~\bibnamefont {Rothuizen}}, \bibinfo {author}
  {\bibfnamefont {M.}~\bibnamefont {Despont}}, \bibinfo {author} {\bibfnamefont
  {P.}~\bibnamefont {Vettiger}}, \bibinfo {author} {\bibfnamefont
  {F.}~\bibnamefont {Battiston}},  \emph {et~al.},\ }\href@noop {} {\bibfield
  {journal} {\bibinfo  {journal} {Ultramicroscopy}\ }\textbf {\bibinfo {volume}
  {{82}}},\ \bibinfo {pages} {1} (\bibinfo {year} {2000})}\BibitemShut
  {NoStop}%
\bibitem [{\citenamefont {Furano}\ \emph {et~al.}(2013)\citenamefont {Furano},
  \citenamefont {Jansen},\ and\ \citenamefont {Menicucci}}]{furano2013review}%
  \BibitemOpen
  \bibfield  {author} {\bibinfo {author} {\bibfnamefont {G.}~\bibnamefont
  {Furano}}, \bibinfo {author} {\bibfnamefont {R.}~\bibnamefont {Jansen}}, \
  and\ \bibinfo {author} {\bibfnamefont {A.}~\bibnamefont {Menicucci}},\
  }\href@noop {} {\bibfield  {journal} {\bibinfo  {journal} {Journal of
  Instrumentation}\ }\textbf {\bibinfo {volume} {{8}}},\ \bibinfo {pages}
  {C02007} (\bibinfo {year} {2013})}\BibitemShut {NoStop}%
\bibitem [{\citenamefont {Pechal}\ \emph {et~al.}(2018)\citenamefont {Pechal},
  \citenamefont {Arrangoiz-Arriola},\ and\ \citenamefont
  {Safavi-Naeini}}]{pechal2018superconducting}%
  \BibitemOpen
  \bibfield  {author} {\bibinfo {author} {\bibfnamefont {M.}~\bibnamefont
  {Pechal}}, \bibinfo {author} {\bibfnamefont {P.}~\bibnamefont
  {Arrangoiz-Arriola}}, \ and\ \bibinfo {author} {\bibfnamefont {A.~H.}\
  \bibnamefont {Safavi-Naeini}},\ }\href@noop {} {\bibfield  {journal}
  {\bibinfo  {journal} {Quantum Science and Technology}\ }\textbf {\bibinfo
  {volume} {4}},\ \bibinfo {pages} {015006} (\bibinfo {year}
  {2018})}\BibitemShut {NoStop}%
\bibitem [{\citenamefont {Bienfait}\ \emph {et~al.}(2019)\citenamefont
  {Bienfait}, \citenamefont {Satzinger}, \citenamefont {Zhong}, \citenamefont
  {Chang}, \citenamefont {Chou}, \citenamefont {Conner}, \citenamefont {Dumur},
  \citenamefont {Grebel}, \citenamefont {Peairs}, \citenamefont {Povey} \emph
  {et~al.}}]{bienfait2019phonon}%
  \BibitemOpen
  \bibfield  {author} {\bibinfo {author} {\bibfnamefont {A.}~\bibnamefont
  {Bienfait}}, \bibinfo {author} {\bibfnamefont {K.~J.}\ \bibnamefont
  {Satzinger}}, \bibinfo {author} {\bibfnamefont {Y.}~\bibnamefont {Zhong}},
  \bibinfo {author} {\bibfnamefont {H.-S.}\ \bibnamefont {Chang}}, \bibinfo
  {author} {\bibfnamefont {M.-H.}\ \bibnamefont {Chou}}, \bibinfo {author}
  {\bibfnamefont {C.~R.}\ \bibnamefont {Conner}}, \bibinfo {author}
  {\bibfnamefont {{\'E}.}~\bibnamefont {Dumur}}, \bibinfo {author}
  {\bibfnamefont {J.}~\bibnamefont {Grebel}}, \bibinfo {author} {\bibfnamefont
  {G.~A.}\ \bibnamefont {Peairs}}, \bibinfo {author} {\bibfnamefont {R.~G.}\
  \bibnamefont {Povey}},  \emph {et~al.},\ }\href@noop {} {\bibfield  {journal}
  {\bibinfo  {journal} {Science}\ }\textbf {\bibinfo {volume} {{364}}},\
  \bibinfo {pages} {368} (\bibinfo {year} {2019})}\BibitemShut {NoStop}%
\bibitem [{\citenamefont {Satzinger}\ \emph {et~al.}(2018)\citenamefont
  {Satzinger}, \citenamefont {Zhong}, \citenamefont {Chang}, \citenamefont
  {Peairs}, \citenamefont {Bienfait}, \citenamefont {Chou}, \citenamefont
  {Cleland}, \citenamefont {Conner}, \citenamefont {Dumur}, \citenamefont
  {Grebel} \emph {et~al.}}]{satzinger2018quantum}%
  \BibitemOpen
  \bibfield  {author} {\bibinfo {author} {\bibfnamefont {K.~J.}\ \bibnamefont
  {Satzinger}}, \bibinfo {author} {\bibfnamefont {Y.}~\bibnamefont {Zhong}},
  \bibinfo {author} {\bibfnamefont {H.-S.}\ \bibnamefont {Chang}}, \bibinfo
  {author} {\bibfnamefont {G.~A.}\ \bibnamefont {Peairs}}, \bibinfo {author}
  {\bibfnamefont {A.}~\bibnamefont {Bienfait}}, \bibinfo {author}
  {\bibfnamefont {M.-H.}\ \bibnamefont {Chou}}, \bibinfo {author}
  {\bibfnamefont {A.}~\bibnamefont {Cleland}}, \bibinfo {author} {\bibfnamefont
  {C.~R.}\ \bibnamefont {Conner}}, \bibinfo {author} {\bibfnamefont
  {{\'E}.}~\bibnamefont {Dumur}}, \bibinfo {author} {\bibfnamefont
  {J.}~\bibnamefont {Grebel}},  \emph {et~al.},\ }\href@noop {} {\bibfield
  {journal} {\bibinfo  {journal} {Nature}\ }\textbf {\bibinfo {volume}
  {{563}}},\ \bibinfo {pages} {661} (\bibinfo {year} {2018})}\BibitemShut
  {NoStop}%
\bibitem [{\citenamefont {Shin}\ \emph {et~al.}(2015)\citenamefont {Shin},
  \citenamefont {Cox}, \citenamefont {Jarecki}, \citenamefont {Starbuck},
  \citenamefont {Wang},\ and\ \citenamefont {Rakich}}]{shin2015control}%
  \BibitemOpen
  \bibfield  {author} {\bibinfo {author} {\bibfnamefont {H.}~\bibnamefont
  {Shin}}, \bibinfo {author} {\bibfnamefont {J.~A.}\ \bibnamefont {Cox}},
  \bibinfo {author} {\bibfnamefont {R.}~\bibnamefont {Jarecki}}, \bibinfo
  {author} {\bibfnamefont {A.}~\bibnamefont {Starbuck}}, \bibinfo {author}
  {\bibfnamefont {Z.}~\bibnamefont {Wang}}, \ and\ \bibinfo {author}
  {\bibfnamefont {P.~T.}\ \bibnamefont {Rakich}},\ }\href@noop {} {\bibfield
  {journal} {\bibinfo  {journal} {Nature communications}\ }\textbf {\bibinfo
  {volume} {{6}}},\ \bibinfo {pages} {6427} (\bibinfo {year}
  {2015})}\BibitemShut {NoStop}%
\bibitem [{\citenamefont {Fu}\ \emph {et~al.}(2019)\citenamefont {Fu},
  \citenamefont {Shen}, \citenamefont {Xu}, \citenamefont {Zou}, \citenamefont
  {Cheng}, \citenamefont {Han},\ and\ \citenamefont {Tang}}]{fu2019phononic}%
  \BibitemOpen
  \bibfield  {author} {\bibinfo {author} {\bibfnamefont {W.}~\bibnamefont
  {Fu}}, \bibinfo {author} {\bibfnamefont {Z.}~\bibnamefont {Shen}}, \bibinfo
  {author} {\bibfnamefont {Y.}~\bibnamefont {Xu}}, \bibinfo {author}
  {\bibfnamefont {C.-L.}\ \bibnamefont {Zou}}, \bibinfo {author} {\bibfnamefont
  {R.}~\bibnamefont {Cheng}}, \bibinfo {author} {\bibfnamefont
  {X.}~\bibnamefont {Han}}, \ and\ \bibinfo {author} {\bibfnamefont {H.~X.}\
  \bibnamefont {Tang}},\ }\href@noop {} {\bibfield  {journal} {\bibinfo
  {journal} {Nature communications}\ }\textbf {\bibinfo {volume} {{10}}},\
  \bibinfo {pages} {1} (\bibinfo {year} {2019})}\BibitemShut {NoStop}%
\bibitem [{\citenamefont {Longo}\ \emph {et~al.}(2013)\citenamefont {Longo},
  \citenamefont {Alonso-Sarduy}, \citenamefont {Rio}, \citenamefont {Bizzini},
  \citenamefont {Trampuz}, \citenamefont {Notz}, \citenamefont {Dietler},\ and\
  \citenamefont {Kasas}}]{longo2013rapid}%
  \BibitemOpen
  \bibfield  {author} {\bibinfo {author} {\bibfnamefont {G.}~\bibnamefont
  {Longo}}, \bibinfo {author} {\bibfnamefont {L.}~\bibnamefont
  {Alonso-Sarduy}}, \bibinfo {author} {\bibfnamefont {L.~M.}\ \bibnamefont
  {Rio}}, \bibinfo {author} {\bibfnamefont {A.}~\bibnamefont {Bizzini}},
  \bibinfo {author} {\bibfnamefont {A.}~\bibnamefont {Trampuz}}, \bibinfo
  {author} {\bibfnamefont {J.}~\bibnamefont {Notz}}, \bibinfo {author}
  {\bibfnamefont {G.}~\bibnamefont {Dietler}}, \ and\ \bibinfo {author}
  {\bibfnamefont {S.}~\bibnamefont {Kasas}},\ }\href@noop {} {\bibfield
  {journal} {\bibinfo  {journal} {Nature nanotechnology}\ }\textbf {\bibinfo
  {volume} {{8}}},\ \bibinfo {pages} {522} (\bibinfo {year}
  {2013})}\BibitemShut {NoStop}%
\bibitem [{\citenamefont {Kurosu}\ \emph {et~al.}(2018)\citenamefont {Kurosu},
  \citenamefont {Hatanaka}, \citenamefont {Onomitsu},\ and\ \citenamefont
  {Yamaguchi}}]{kurosu_-chip_2018}%
  \BibitemOpen
  \bibfield  {author} {\bibinfo {author} {\bibfnamefont {M.}~\bibnamefont
  {Kurosu}}, \bibinfo {author} {\bibfnamefont {D.}~\bibnamefont {Hatanaka}},
  \bibinfo {author} {\bibfnamefont {K.}~\bibnamefont {Onomitsu}}, \ and\
  \bibinfo {author} {\bibfnamefont {H.}~\bibnamefont {Yamaguchi}},\ }\href@noop
  {} {\bibfield  {journal} {\bibinfo  {journal} {Nature Communications}\
  }\textbf {\bibinfo {volume} {9}},\ \bibinfo {pages} {1331} (\bibinfo {year}
  {2018})}\BibitemShut {NoStop}%
\bibitem [{Sup()}]{Supp}%
  \BibitemOpen
  \href@noop {} {}\bibinfo {note} {See Supplemental Material for further
  information}\BibitemShut {NoStop}%
\bibitem [{\citenamefont {Wenzler}\ \emph {et~al.}(2014)\citenamefont
  {Wenzler}, \citenamefont {Dunn}, \citenamefont {Toffoli},\ and\ \citenamefont
  {Mohanty}}]{wenzler2014nanomechanical}%
  \BibitemOpen
  \bibfield  {author} {\bibinfo {author} {\bibfnamefont {J.-S.}\ \bibnamefont
  {Wenzler}}, \bibinfo {author} {\bibfnamefont {T.}~\bibnamefont {Dunn}},
  \bibinfo {author} {\bibfnamefont {T.}~\bibnamefont {Toffoli}}, \ and\
  \bibinfo {author} {\bibfnamefont {P.}~\bibnamefont {Mohanty}},\ }\href@noop
  {} {\bibfield  {journal} {\bibinfo  {journal} {Nano letters}\ }\textbf
  {\bibinfo {volume} {14}},\ \bibinfo {pages} {89} (\bibinfo {year}
  {2014})}\BibitemShut {NoStop}%
\bibitem [{\citenamefont {Lee}\ and\ \citenamefont
  {Nguyen}(2004)}]{lee2004mechanically}%
  \BibitemOpen
  \bibfield  {author} {\bibinfo {author} {\bibfnamefont {S.}~\bibnamefont
  {Lee}}\ and\ \bibinfo {author} {\bibfnamefont {C.-C.}\ \bibnamefont
  {Nguyen}},\ }in\ \href@noop {} {\emph {\bibinfo {booktitle} {Proceedings of
  the 2004 IEEE International Frequency Control Symposium and Exposition,
  2004.}}}\ (\bibinfo {organization} {IEEE},\ \bibinfo {year} {2004})\ pp.\
  \bibinfo {pages} {144--150}\BibitemShut {NoStop}%
\bibitem [{\citenamefont {Demirci}\ and\ \citenamefont
  {Nguyen}(2006)}]{demirci2006mechanically}%
  \BibitemOpen
  \bibfield  {author} {\bibinfo {author} {\bibfnamefont {M.~U.}\ \bibnamefont
  {Demirci}}\ and\ \bibinfo {author} {\bibfnamefont {C.~T.-C.}\ \bibnamefont
  {Nguyen}},\ }\href@noop {} {\bibfield  {journal} {\bibinfo  {journal}
  {Journal of Microelectromechanical Systems}\ }\textbf {\bibinfo {volume}
  {15}},\ \bibinfo {pages} {1419} (\bibinfo {year} {2006})}\BibitemShut
  {NoStop}%
\bibitem [{\citenamefont {Greywall}\ and\ \citenamefont
  {Busch}(2002)}]{greywall2002coupled}%
  \BibitemOpen
  \bibfield  {author} {\bibinfo {author} {\bibfnamefont {D.~S.}\ \bibnamefont
  {Greywall}}\ and\ \bibinfo {author} {\bibfnamefont {P.~A.}\ \bibnamefont
  {Busch}},\ }\href@noop {} {\bibfield  {journal} {\bibinfo  {journal} {Journal
  of Micromechanics and Microengineering}\ }\textbf {\bibinfo {volume} {12}},\
  \bibinfo {pages} {925} (\bibinfo {year} {2002})}\BibitemShut {NoStop}%
\bibitem [{\citenamefont {Doster}\ \emph {et~al.}(2019)\citenamefont {Doster},
  \citenamefont {H{\"o}nl}, \citenamefont {Lorenz}, \citenamefont
  {Paulitschke},\ and\ \citenamefont {Weig}}]{doster2019collective}%
  \BibitemOpen
  \bibfield  {author} {\bibinfo {author} {\bibfnamefont {J.}~\bibnamefont
  {Doster}}, \bibinfo {author} {\bibfnamefont {S.}~\bibnamefont {H{\"o}nl}},
  \bibinfo {author} {\bibfnamefont {H.}~\bibnamefont {Lorenz}}, \bibinfo
  {author} {\bibfnamefont {P.}~\bibnamefont {Paulitschke}}, \ and\ \bibinfo
  {author} {\bibfnamefont {E.~M.}\ \bibnamefont {Weig}},\ }\href@noop {}
  {\bibfield  {journal} {\bibinfo  {journal} {Nature communications}\ }\textbf
  {\bibinfo {volume} {10}},\ \bibinfo {pages} {1} (\bibinfo {year}
  {2019})}\BibitemShut {NoStop}%
\bibitem [{\citenamefont {Romero}\ \emph {et~al.}(2019)\citenamefont {Romero},
  \citenamefont {Kalra}, \citenamefont {Mauranyapin}, \citenamefont {Baker},
  \citenamefont {Meng},\ and\ \citenamefont {Bowen}}]{romero_propagation_2019}%
  \BibitemOpen
  \bibfield  {author} {\bibinfo {author} {\bibfnamefont {E.}~\bibnamefont
  {Romero}}, \bibinfo {author} {\bibfnamefont {R.}~\bibnamefont {Kalra}},
  \bibinfo {author} {\bibfnamefont {N.}~\bibnamefont {Mauranyapin}}, \bibinfo
  {author} {\bibfnamefont {C.}~\bibnamefont {Baker}}, \bibinfo {author}
  {\bibfnamefont {C.}~\bibnamefont {Meng}}, \ and\ \bibinfo {author}
  {\bibfnamefont {W.}~\bibnamefont {Bowen}},\ }\href@noop {} {\bibfield
  {journal} {\bibinfo  {journal} {Phys. Rev. Applied}\ }\textbf {\bibinfo
  {volume} {{11}}},\ \bibinfo {pages} {064035} (\bibinfo {year}
  {2019})}\BibitemShut {NoStop}%
\bibitem [{\citenamefont {Romero}\ \emph {et~al.}(2020)\citenamefont {Romero},
  \citenamefont {Valenzuela}, \citenamefont {Kermany}, \citenamefont
  {Sementilli}, \citenamefont {Iacopi},\ and\ \citenamefont
  {Bowen}}]{romero_engineering_2020}%
  \BibitemOpen
  \bibfield  {author} {\bibinfo {author} {\bibfnamefont {E.}~\bibnamefont
  {Romero}}, \bibinfo {author} {\bibfnamefont {V.~M.}\ \bibnamefont
  {Valenzuela}}, \bibinfo {author} {\bibfnamefont {A.~R.}\ \bibnamefont
  {Kermany}}, \bibinfo {author} {\bibfnamefont {L.}~\bibnamefont {Sementilli}},
  \bibinfo {author} {\bibfnamefont {F.}~\bibnamefont {Iacopi}}, \ and\ \bibinfo
  {author} {\bibfnamefont {W.~P.}\ \bibnamefont {Bowen}},\ }\href@noop {}
  {\bibfield  {journal} {\bibinfo  {journal} {Phys. Rev. Applied}\ }\textbf
  {\bibinfo {volume} {{13}}},\ \bibinfo {pages} {044007} (\bibinfo {year}
  {2020})},\ \bibinfo {note} {publisher: American Physical Society}\BibitemShut
  {NoStop}%
\bibitem [{\citenamefont {Hatanaka}\ \emph {et~al.}(2014)\citenamefont
  {Hatanaka}, \citenamefont {Mahboob}, \citenamefont {Onomitsu},\ and\
  \citenamefont {Yamaguchi}}]{hatanaka2014phonon}%
  \BibitemOpen
  \bibfield  {author} {\bibinfo {author} {\bibfnamefont {D.}~\bibnamefont
  {Hatanaka}}, \bibinfo {author} {\bibfnamefont {I.}~\bibnamefont {Mahboob}},
  \bibinfo {author} {\bibfnamefont {K.}~\bibnamefont {Onomitsu}}, \ and\
  \bibinfo {author} {\bibfnamefont {H.}~\bibnamefont {Yamaguchi}},\ }\href@noop
  {} {\bibfield  {journal} {\bibinfo  {journal} {Nature nanotechnology}\
  }\textbf {\bibinfo {volume} {9}},\ \bibinfo {pages} {520} (\bibinfo {year}
  {2014})}\BibitemShut {NoStop}%
\bibitem [{\citenamefont {Cha}\ \emph {et~al.}(2018)\citenamefont {Cha},
  \citenamefont {Kim},\ and\ \citenamefont {Daraio}}]{cha_experimental_2018}%
  \BibitemOpen
  \bibfield  {author} {\bibinfo {author} {\bibfnamefont {J.}~\bibnamefont
  {Cha}}, \bibinfo {author} {\bibfnamefont {K.~W.}\ \bibnamefont {Kim}}, \ and\
  \bibinfo {author} {\bibfnamefont {C.}~\bibnamefont {Daraio}},\ }\href@noop {}
  {\bibfield  {journal} {\bibinfo  {journal} {Nature}\ }\textbf {\bibinfo
  {volume} {{564}}},\ \bibinfo {pages} {229} (\bibinfo {year}
  {2018})}\BibitemShut {NoStop}%
\bibitem [{\citenamefont {Kurosu}\ \emph {et~al.}(2020)\citenamefont {Kurosu},
  \citenamefont {Hatanaka},\ and\ \citenamefont
  {Yamaguchi}}]{kurosu_mechanical_2020}%
  \BibitemOpen
  \bibfield  {author} {\bibinfo {author} {\bibfnamefont {M.}~\bibnamefont
  {Kurosu}}, \bibinfo {author} {\bibfnamefont {D.}~\bibnamefont {Hatanaka}}, \
  and\ \bibinfo {author} {\bibfnamefont {H.}~\bibnamefont {Yamaguchi}},\
  }\href@noop {} {\bibfield  {journal} {\bibinfo  {journal} {Phys. Rev.
  Applied}\ }\textbf {\bibinfo {volume} {{13}}},\ \bibinfo {pages} {014056}
  (\bibinfo {year} {2020})}\BibitemShut {NoStop}%
\bibitem [{\citenamefont {Cha}\ and\ \citenamefont
  {Daraio}(2018)}]{cha_electrical_2018}%
  \BibitemOpen
  \bibfield  {author} {\bibinfo {author} {\bibfnamefont {J.}~\bibnamefont
  {Cha}}\ and\ \bibinfo {author} {\bibfnamefont {C.}~\bibnamefont {Daraio}},\
  }\href@noop {} {\bibfield  {journal} {\bibinfo  {journal} {Nature Nanotech}\
  }\textbf {\bibinfo {volume} {{13}}},\ \bibinfo {pages} {1016} (\bibinfo
  {year} {2018})}\BibitemShut {NoStop}%
\bibitem [{\citenamefont {Mauranyapin}\ \emph {et~al.}(2017)\citenamefont
  {Mauranyapin}, \citenamefont {Madsen}, \citenamefont {Taylor}, \citenamefont
  {Waleed},\ and\ \citenamefont {Bowen}}]{mauranyapin_evanescent_2017}%
  \BibitemOpen
  \bibfield  {author} {\bibinfo {author} {\bibfnamefont {N.}~\bibnamefont
  {Mauranyapin}}, \bibinfo {author} {\bibfnamefont {L.}~\bibnamefont {Madsen}},
  \bibinfo {author} {\bibfnamefont {M.}~\bibnamefont {Taylor}}, \bibinfo
  {author} {\bibfnamefont {M.}~\bibnamefont {Waleed}}, \ and\ \bibinfo {author}
  {\bibfnamefont {W.}~\bibnamefont {Bowen}},\ }\href@noop {} {\bibfield
  {journal} {\bibinfo  {journal} {Nature Photonics}\ }\textbf {\bibinfo
  {volume} {{11}}},\ \bibinfo {pages} {477} (\bibinfo {year}
  {2017})}\BibitemShut {NoStop}%
\bibitem [{\citenamefont {Stahlhofen}\ and\ \citenamefont
  {Nimtz}(2006)}]{stahlhofen2006evanescent}%
  \BibitemOpen
  \bibfield  {author} {\bibinfo {author} {\bibfnamefont {A.}~\bibnamefont
  {Stahlhofen}}\ and\ \bibinfo {author} {\bibfnamefont {G.}~\bibnamefont
  {Nimtz}},\ }\href@noop {} {\bibfield  {journal} {\bibinfo  {journal} {EPL
  (Europhysics Letters)}\ }\textbf {\bibinfo {volume} {76}},\ \bibinfo {pages}
  {189} (\bibinfo {year} {2006})}\BibitemShut {NoStop}%
\bibitem [{\citenamefont {Luo}\ \emph {et~al.}(2014)\citenamefont {Luo},
  \citenamefont {Ophir}, \citenamefont {Chen}, \citenamefont {Gabrielli},
  \citenamefont {Poitras}, \citenamefont {Bergmen},\ and\ \citenamefont
  {Lipson}}]{luo2014wdm}%
  \BibitemOpen
  \bibfield  {author} {\bibinfo {author} {\bibfnamefont {L.-W.}\ \bibnamefont
  {Luo}}, \bibinfo {author} {\bibfnamefont {N.}~\bibnamefont {Ophir}}, \bibinfo
  {author} {\bibfnamefont {C.~P.}\ \bibnamefont {Chen}}, \bibinfo {author}
  {\bibfnamefont {L.~H.}\ \bibnamefont {Gabrielli}}, \bibinfo {author}
  {\bibfnamefont {C.~B.}\ \bibnamefont {Poitras}}, \bibinfo {author}
  {\bibfnamefont {K.}~\bibnamefont {Bergmen}}, \ and\ \bibinfo {author}
  {\bibfnamefont {M.}~\bibnamefont {Lipson}},\ }\href@noop {} {\bibfield
  {journal} {\bibinfo  {journal} {Nature communications}\ }\textbf {\bibinfo
  {volume} {{5}}},\ \bibinfo {pages} {1} (\bibinfo {year} {2014})}\BibitemShut
  {NoStop}%
\bibitem [{\citenamefont {Fang}\ \emph {et~al.}(2016)\citenamefont {Fang},
  \citenamefont {Matheny}, \citenamefont {Luan},\ and\ \citenamefont
  {Painter}}]{fang2016optical}%
  \BibitemOpen
  \bibfield  {author} {\bibinfo {author} {\bibfnamefont {K.}~\bibnamefont
  {Fang}}, \bibinfo {author} {\bibfnamefont {M.~H.}\ \bibnamefont {Matheny}},
  \bibinfo {author} {\bibfnamefont {X.}~\bibnamefont {Luan}}, \ and\ \bibinfo
  {author} {\bibfnamefont {O.}~\bibnamefont {Painter}},\ }\href@noop {}
  {\bibfield  {journal} {\bibinfo  {journal} {Nature Photonics}\ }\textbf
  {\bibinfo {volume} {10}},\ \bibinfo {pages} {489} (\bibinfo {year}
  {2016})}\BibitemShut {NoStop}%
\bibitem [{\citenamefont {Haus}\ and\ \citenamefont
  {Huang}(1991)}]{haus1991coupled}%
  \BibitemOpen
  \bibfield  {author} {\bibinfo {author} {\bibfnamefont {H.~A.}\ \bibnamefont
  {Haus}}\ and\ \bibinfo {author} {\bibfnamefont {W.}~\bibnamefont {Huang}},\
  }\href@noop {} {\bibfield  {journal} {\bibinfo  {journal} {Proceedings of the
  IEEE}\ }\textbf {\bibinfo {volume} {{79}}},\ \bibinfo {pages} {1505}
  (\bibinfo {year} {1991})}\BibitemShut {NoStop}%
\bibitem [{\citenamefont {Tian}\ \emph {et~al.}(2017)\citenamefont {Tian},
  \citenamefont {Liu}, \citenamefont {Xiao}, \citenamefont {Zhao},
  \citenamefont {Liu}, \citenamefont {Yang}, \citenamefont {Ding},
  \citenamefont {Zhang},\ and\ \citenamefont {Yang}}]{tian2017experimental}%
  \BibitemOpen
  \bibfield  {author} {\bibinfo {author} {\bibfnamefont {Y.}~\bibnamefont
  {Tian}}, \bibinfo {author} {\bibfnamefont {Z.}~\bibnamefont {Liu}}, \bibinfo
  {author} {\bibfnamefont {H.}~\bibnamefont {Xiao}}, \bibinfo {author}
  {\bibfnamefont {G.}~\bibnamefont {Zhao}}, \bibinfo {author} {\bibfnamefont
  {G.}~\bibnamefont {Liu}}, \bibinfo {author} {\bibfnamefont {J.}~\bibnamefont
  {Yang}}, \bibinfo {author} {\bibfnamefont {J.}~\bibnamefont {Ding}}, \bibinfo
  {author} {\bibfnamefont {L.}~\bibnamefont {Zhang}}, \ and\ \bibinfo {author}
  {\bibfnamefont {L.}~\bibnamefont {Yang}},\ }\href@noop {} {\bibfield
  {journal} {\bibinfo  {journal} {Scientific reports}\ }\textbf {\bibinfo
  {volume} {{7}}},\ \bibinfo {pages} {1} (\bibinfo {year} {2017})}\BibitemShut
  {NoStop}%
\bibitem [{\citenamefont {Gabrielli}\ \emph {et~al.}(2009)\citenamefont
  {Gabrielli}, \citenamefont {Cardenas}, \citenamefont {Poitras},\ and\
  \citenamefont {Lipson}}]{gabrielli2009silicon}%
  \BibitemOpen
  \bibfield  {author} {\bibinfo {author} {\bibfnamefont {L.~H.}\ \bibnamefont
  {Gabrielli}}, \bibinfo {author} {\bibfnamefont {J.}~\bibnamefont {Cardenas}},
  \bibinfo {author} {\bibfnamefont {C.~B.}\ \bibnamefont {Poitras}}, \ and\
  \bibinfo {author} {\bibfnamefont {M.}~\bibnamefont {Lipson}},\ }\href@noop {}
  {\bibfield  {journal} {\bibinfo  {journal} {Nature photonics}\ }\textbf
  {\bibinfo {volume} {{3}}},\ \bibinfo {pages} {461} (\bibinfo {year}
  {2009})}\BibitemShut {NoStop}%
\bibitem [{\citenamefont {Qiang}\ \emph {et~al.}(2007)\citenamefont {Qiang},
  \citenamefont {Zhou},\ and\ \citenamefont {Soref}}]{qiang2007optical}%
  \BibitemOpen
  \bibfield  {author} {\bibinfo {author} {\bibfnamefont {Z.}~\bibnamefont
  {Qiang}}, \bibinfo {author} {\bibfnamefont {W.}~\bibnamefont {Zhou}}, \ and\
  \bibinfo {author} {\bibfnamefont {R.~A.}\ \bibnamefont {Soref}},\ }\href@noop
  {} {\bibfield  {journal} {\bibinfo  {journal} {Optics express}\ }\textbf
  {\bibinfo {volume} {{15}}},\ \bibinfo {pages} {1823} (\bibinfo {year}
  {2007})}\BibitemShut {NoStop}%
\bibitem [{\citenamefont {Bekker}\ \emph {et~al.}(2018)\citenamefont {Bekker},
  \citenamefont {Baker}, \citenamefont {Kalra}, \citenamefont {Cheng},
  \citenamefont {Li}, \citenamefont {Prakash},\ and\ \citenamefont
  {Bowen}}]{bekker2018free}%
  \BibitemOpen
  \bibfield  {author} {\bibinfo {author} {\bibfnamefont {C.}~\bibnamefont
  {Bekker}}, \bibinfo {author} {\bibfnamefont {C.~G.}\ \bibnamefont {Baker}},
  \bibinfo {author} {\bibfnamefont {R.}~\bibnamefont {Kalra}}, \bibinfo
  {author} {\bibfnamefont {H.-H.}\ \bibnamefont {Cheng}}, \bibinfo {author}
  {\bibfnamefont {B.-B.}\ \bibnamefont {Li}}, \bibinfo {author} {\bibfnamefont
  {V.}~\bibnamefont {Prakash}}, \ and\ \bibinfo {author} {\bibfnamefont
  {W.~P.}\ \bibnamefont {Bowen}},\ }\href@noop {} {\bibfield  {journal}
  {\bibinfo  {journal} {Optics express}\ }\textbf {\bibinfo {volume} {{26}}},\
  \bibinfo {pages} {33649} (\bibinfo {year} {2018})}\BibitemShut {NoStop}%
\bibitem [{\citenamefont {Narayanamurti}\ \emph {et~al.}(1979)\citenamefont
  {Narayanamurti}, \citenamefont {St{\"o}rmer}, \citenamefont {Chin},
  \citenamefont {Gossard},\ and\ \citenamefont
  {Wiegmann}}]{narayanamurti1979selective}%
  \BibitemOpen
  \bibfield  {author} {\bibinfo {author} {\bibfnamefont {V.}~\bibnamefont
  {Narayanamurti}}, \bibinfo {author} {\bibfnamefont {H.}~\bibnamefont
  {St{\"o}rmer}}, \bibinfo {author} {\bibfnamefont {M.}~\bibnamefont {Chin}},
  \bibinfo {author} {\bibfnamefont {A.}~\bibnamefont {Gossard}}, \ and\
  \bibinfo {author} {\bibfnamefont {W.}~\bibnamefont {Wiegmann}},\ }\href@noop
  {} {\bibfield  {journal} {\bibinfo  {journal} {Physical Review Letters}\
  }\textbf {\bibinfo {volume} {43}},\ \bibinfo {pages} {2012} (\bibinfo {year}
  {1979})}\BibitemShut {NoStop}%
\bibitem [{\citenamefont {Catalini}\ \emph {et~al.}(2020)\citenamefont
  {Catalini}, \citenamefont {Tsaturyan},\ and\ \citenamefont
  {Schliesser}}]{catalini2020soft}%
  \BibitemOpen
  \bibfield  {author} {\bibinfo {author} {\bibfnamefont {L.}~\bibnamefont
  {Catalini}}, \bibinfo {author} {\bibfnamefont {Y.}~\bibnamefont {Tsaturyan}},
  \ and\ \bibinfo {author} {\bibfnamefont {A.}~\bibnamefont {Schliesser}},\
  }\href@noop {} {\bibfield  {journal} {\bibinfo  {journal} {Physical Review
  Applied}\ }\textbf {\bibinfo {volume} {14}},\ \bibinfo {pages} {014041}
  (\bibinfo {year} {2020})}\BibitemShut {NoStop}%
\bibitem [{\citenamefont {Taylor}\ \emph {et~al.}(2012)\citenamefont {Taylor},
  \citenamefont {Szorkovszky}, \citenamefont {Knittel}, \citenamefont {Lee},
  \citenamefont {McRae},\ and\ \citenamefont {Bowen}}]{taylor2012cavity}%
  \BibitemOpen
  \bibfield  {author} {\bibinfo {author} {\bibfnamefont {M.~A.}\ \bibnamefont
  {Taylor}}, \bibinfo {author} {\bibfnamefont {A.}~\bibnamefont {Szorkovszky}},
  \bibinfo {author} {\bibfnamefont {J.}~\bibnamefont {Knittel}}, \bibinfo
  {author} {\bibfnamefont {K.~H.}\ \bibnamefont {Lee}}, \bibinfo {author}
  {\bibfnamefont {T.~G.}\ \bibnamefont {McRae}}, \ and\ \bibinfo {author}
  {\bibfnamefont {W.~P.}\ \bibnamefont {Bowen}},\ }\href@noop {} {\bibfield
  {journal} {\bibinfo  {journal} {Optics express}\ }\textbf {\bibinfo {volume}
  {{20}}},\ \bibinfo {pages} {12742} (\bibinfo {year} {2012})}\BibitemShut
  {NoStop}%
\bibitem [{\citenamefont {Fedorov}\ \emph {et~al.}(2019)\citenamefont
  {Fedorov}, \citenamefont {Engelsen}, \citenamefont {Ghadimi}, \citenamefont
  {Bereyhi}, \citenamefont {Schilling}, \citenamefont {Wilson},\ and\
  \citenamefont {Kippenberg}}]{fedorov2019generalized}%
  \BibitemOpen
  \bibfield  {author} {\bibinfo {author} {\bibfnamefont {S.~A.}\ \bibnamefont
  {Fedorov}}, \bibinfo {author} {\bibfnamefont {N.~J.}\ \bibnamefont
  {Engelsen}}, \bibinfo {author} {\bibfnamefont {A.~H.}\ \bibnamefont
  {Ghadimi}}, \bibinfo {author} {\bibfnamefont {M.~J.}\ \bibnamefont
  {Bereyhi}}, \bibinfo {author} {\bibfnamefont {R.}~\bibnamefont {Schilling}},
  \bibinfo {author} {\bibfnamefont {D.~J.}\ \bibnamefont {Wilson}}, \ and\
  \bibinfo {author} {\bibfnamefont {T.~J.}\ \bibnamefont {Kippenberg}},\
  }\href@noop {} {\bibfield  {journal} {\bibinfo  {journal} {Physical Review
  B}\ }\textbf {\bibinfo {volume} {{99}}},\ \bibinfo {pages} {054107} (\bibinfo
  {year} {(2019)})}\BibitemShut {NoStop}%
\bibitem [{\citenamefont {Naesby}\ \emph {et~al.}(2017)\citenamefont {Naesby},
  \citenamefont {Naserbakht},\ and\ \citenamefont
  {Dantan}}]{naesby_effects_2017}%
  \BibitemOpen
  \bibfield  {author} {\bibinfo {author} {\bibfnamefont {A.}~\bibnamefont
  {Naesby}}, \bibinfo {author} {\bibfnamefont {S.}~\bibnamefont {Naserbakht}},
  \ and\ \bibinfo {author} {\bibfnamefont {A.}~\bibnamefont {Dantan}},\
  }\href@noop {} {\bibfield  {journal} {\bibinfo  {journal} {Applied Physics
  Letters}\ }\textbf {\bibinfo {volume} {{111}}},\ \bibinfo {pages} {201103}
  (\bibinfo {year} {(2017)})},\ \bibinfo {note} {publisher: American Institute
  of Physics}\BibitemShut {NoStop}%
\end{thebibliography}%

\end{document}